\documentclass[aps,prd,twocolumn,groupedaddress,nofootinbib,
superscriptaddress,preprintnumbers]{revtex4-1}
\usepackage{epsfig,graphicx}
\usepackage{dcolumn}
\usepackage{bm}
\usepackage{amsfonts}
\usepackage{amssymb}
\usepackage{amsmath}

\usepackage{color}

\usepackage[hidelinks]{hyperref}

\usepackage{soul}

\renewcommand{\ss}{\scriptscriptstyle}

\begin{document}

\title{Einstein-Gauss-Bonnet gravity in 4-dimensional space-time} 

\author{Dra\v{z}en Glavan}
\email[]{drazen.glavan@uclouvain.be}
\affiliation{Centre for Cosmology, Particle Physics and Phenomenology (CP3),
Universit\'{e} catholique de Louvain, 
Chemin du Cyclotron 2, 1348 Louvain-la-Neuve, Belgium}

\author{Chunshan Lin}
\email[]{chunshan.lin@fuw.edu.pl}
\affiliation{Institute of Theoretical Physics, Faculty of Physics,
University of Warsaw, Pasteura 5, 02-093 Warsaw, Poland}

\preprint{CP3-19-24}

\begin{abstract}

In this {\it Letter} we present a general covariant modified theory of 
gravity in~$D\!=\!4$ space-time dimensions which propagates only 
the massless graviton {\it and} bypasses the Lovelock's theorem.
The theory we present is formulated in~$D\!>\!4$ dimensions and
its action consists of the Einstein-Hilbert term with a cosmological constant,
and the Gauss-Bonnet term multiplied by a  factor~$1/(D\!-\!4)$.
The four-dimensional theory is defined as the limit~$D\!\to\!4$. 
In this 
singular limit the Gauss-Bonnet invariant gives rise to 
non-trivial contributions to gravitational dynamics, while preserving the
number of graviton degrees of freedom and being free from Ostrogradsky
instability. We report several 
appealing new predictions of this theory, including the corrections to 
the dispersion relation of cosmological tensor and scalar modes, 
singularity resolution for spherically symmetric solutions, and others.

\end{abstract}

\maketitle

{\bf Introduction.}
~According to the Lovelock's theorem~\cite{Lovelock:1971yv,Lovelock:1972vz,
Lanczos:1938sf}, Einstein's general relativity with the cosmological constant 
is the unique theory of gravity if we assume: ($i$) the space time is 3+1 dimensional,
($ii$) diffeomorphism invariance, ($iii$) metricity,  and ($iv$) second order equations of motion. In 
this {\it Letter} we demonstrate a way to bypass the conclusions of Lovelock's theorem, and 
present a model respecting all the assumptions ($i$-$iv$), but nevertheless 
exhibiting modified dynamics. 

It is believed that the most general theory in four dimensional space-time consists of the Einstein-Hilbert action and a cosmological constant,
\begin{equation}
S_{\rm \ss EH}[g_{\mu\nu}]
	= \int \! d^{D\!}x \, \sqrt{-g}
	\left[\frac{M_{\ss \rm P}^2}{2}R-\Lambda_0\right],
\end{equation}
where $D\!=\!4$.  This theory contains two parameters
-- the reduced Planck mass~$M_{\rm \ss P}$ characterizing the gravitational 
coupling strength, and the (bare) cosmological constant~$\Lambda_0$
playing the role of vacuum energy.

In higher dimensions, however, there are more terms -- higher order Lovelock 
invariants -- satisfying conditions ($ii$-$iv$). First such term appears in five dimensions,
\begin{equation}
S_{ \rm \ss GB}[g_{\mu\nu}] 
	= \int\! d^{D\!}x \, \sqrt{-g} \, \alpha \, \mathcal{G} \, ,
\end{equation}
where~$\alpha$ is a dimensionless coupling constant and~$\mathcal{G}$ is the 
Gauss-Bonnet invariant,~$\mathcal{G} \!=\! 
	{R^{\mu\nu}}_{\rho\sigma} {R^{\rho\sigma}}_{\mu\nu}
	\!-\! 4 {R^\mu}_\nu {R^\nu}_\mu \!+\! R^2 \!=\! 
	6 {R^{\mu\nu}}_{[\mu\nu} {R^{\rho\sigma}}_{\rho\sigma]}$.
In~$D\!=\!4$ the Gauss-Bonnet invariant is a total derivative, and hence does not 
contribute to the gravitational dynamics. This is exhibited by its contribution
to Einstein's equation,
\begin{eqnarray}\label{eomgb}
&&
 \frac{ g_{\nu \rho}}{\sqrt{-g}}
	\frac{\delta S_{\rm \ss GB} }{\delta g_{\mu \rho}}
		= 15\alpha \, {\delta^{\mu}}_{[\nu} {R^{\rho\sigma}}_{\rho\sigma} 
		{R^{\alpha\beta}}_{\alpha\beta]} 
		=
		- 2 {R^{\mu\alpha}}_{\rho\sigma} {R^{\rho\sigma}}_{\nu\alpha}
\nonumber
\\
&& \hspace{0.5cm}
	+ 4 {R^{\mu \alpha}}_{\nu \beta} {R^\beta}_\alpha
	+ 4 {R^\mu}_\alpha {R^\alpha}_\nu
	- 2 R {R^\mu}_\nu
	+ \frac{1}{2} \mathcal{G} \delta^{\mu}_{\nu} \, .
\end{eqnarray}
being anti-symmetrized over five indices, and vanishing identically 
in~$D\!=\!4$, but not in~$D\!\ge\!5$. 
An explicit manifestation of this can be seen by taking the trace of~(\ref{eomgb}),
\begin{equation}\label{trace}
 \frac{ g_{\mu\nu}}{\sqrt{-g}}
	\frac{\delta S_{\rm \ss GB} }{\delta g_{\mu \nu}}
		= (D\!-\!4) \times \frac{\alpha}{2} \mathcal{G} \, ,
\end{equation}
which is proportional to a vanishing factor $(D\!-\!4)$ in four 
space-time dimensions.  One might wonder whether this feature is specific just to the trace 
equation or whether it is a general feature of Einstein's equation. 
This question was addressed previously in the literature 
in~\cite{Mardones:1990qc,Torii:2008ru} with the conclusion that 
the Gauss-Bonnet term contribution to all the components of Einstein's equation 
are in fact proportional to~$(D\!-\!4)$, regardless of the space-time symmetries. For instance, for an even dimensional space-time with $D>4$, we have the Einstein-Lovelock equation written in terms of differential form \cite{Mardones:1990qc} 
\begin{eqnarray}\label{eomindf}
\sum_{p=0}^{\frac{D}{2}-1}\alpha_p\left(D \!-\! 2p\right)\epsilon_{a_1...a_D} R^{a_1,a_2}\wedge...\wedge R^{a_{2p-1},a_{2p}}\nonumber\\
\wedge e^{a_{2p+1}}\wedge...\wedge e^{a_{D-1}}=0.
\end{eqnarray}
We $e^a$ is the  {\it vielbein},  and we obtain the factor of $(D\!-\!4)$ for the Gauss-Bonnet term where $p=2$.  Noted that the space-time indices are suppressed  in the above equation, and there is one less  $e^a$ for the odd dimensional space-time. This proportionality to
$(D\!-\!4)$ has also been observed in the 
dynamical equation of motion for graviton in the ADM $D\!=\!d\!+\!1$ decomposition analysis \cite{Torii:2008ru}. 

The idea we investigate in this $Letter$ is the following. 
What if we rescale the coupling constant,
\begin{equation}
\alpha \to \alpha/(D\!-\!4) \, ,
\label{coupling}
\end{equation}
of the Gauss-Bonnet term, 
and then consider the limit~$D\!\rightarrow\!4$? This idea is reminiscent 
of the way in which finite terms are generated by dimensional regularization
in quantum field theory, after the divergences are absorbed by counterterms.
It is  particularly similar to the way in which the
conformal (trace) anomaly arises in quantum field theory in curved 
space-times~\cite{Brown:1976wc}.
However, contrary to dimensional regularization, here there are no divergent 
contributions that need to be subtracted, but rather the singular 
coefficient is introduced to extract a finite contribution from the Gauss-Bonnet term. 
Therefore, we consider this 
prescription to define a {\it classical  theory of gravity}.

Furthermore, what distinguishes this theory from the conformal anomaly 
is an attractive feature that the number of degrees of freedom does not change
as $\alpha\!\to\!0$ in any number of dimensions, thus it smoothly connects to 
general relativity, and is free from the Ostrogradsky instability~\cite{Woodard:2015zca}. 
The same cannot be said of conformal anomaly which introduces additional degrees of 
freedom due to the introduction of higher derivative terms (but if treated
in the same spirit in which they arise -- perturbatively -- this issue can
be circumvented~\cite{Eliezer:1989cr,Glavan:2017srd}).

Therefore, there is no 
obstacle to consider the Gauss-Bonnet
contribution on the same level as the Einstein-Hilbert term. 
Nevertheless, because of Lovelock's theorem, we are prompted to ask whether this 
theory is actually equivalent to Einstein's gravity? As 
will be demonstrated in the remainder of this {\it Letter}, the answer is no. 
\\

{\bf Maximally Symmetric Space-time.}
Let us consider a pure gravity theory given by the 
action~$S\!=\!S_{\ss \rm EH} \!+\! S_{\ss \rm GB}$, {\it i.e.} by
\begin{equation}\label{EGB}
S[g_{\mu\nu}] 
	=\int \! d^{D\!} x \,\sqrt{-g} \,
	\left[\frac{M_{\ss \rm P}^2}{2}R-\Lambda_0+\frac{\alpha}{D\!-\!4} \, \mathcal{G} \right],
\end{equation}
where $\alpha$ is a finite  non-vanishing  dimensionless constant
in $D\!=\!4$. 
Assuming a maximally symmetric 
solution of the theory,  the 
Riemann tensor is given by
by~$M_{\ss \rm P}^2{R^{\mu\nu}}_{\rho\sigma} \!=\! 
	\bigl( \delta^\mu_\rho \delta^\nu_\sigma 
	- \delta^\mu_\sigma \delta^\nu_\rho \bigr)\Lambda/(D\!-\!1)$,
with $\Lambda$ being an effective cosmological
constant. The Gauss-Bonnet contribution
in this case evaluates to 
\begin{equation}
 \frac{ g_{\nu \rho}}{\sqrt{-g}}
\frac{\delta S_{\rm \ss GB} }{\delta g_{\mu \rho}}
	= \frac{\alpha}{D-4}
		\times \frac{(D\!-\!2) (D\!-\!3) (D\!-\!4)}{2(D\!-\!1)M_{\ss \rm P}^4 }
		\times \Lambda^2\delta^\mu_\nu
		\, ,
\label{GB max}
\end{equation}
Note that the divergent factor~$1/(D\!-\!4)$ coming from
the rescaling~(\ref{coupling}) cancels out the vanishing factor~$(D\!-\!4)$
from the variation of the Gauss-Bonnet action.
The same feature is exhibited by all the equations of motion given in the remainder of 
the {\it Letter}. In the limit $D\!\to \!4$, the 
contribution above evaluates to $\alpha\Lambda^2\delta^\mu_\nu/(3M_p^4)$.

There are two branches of solutions for the 
effective cosmological constant,
\begin{equation}
\Lambda_{\pm} \equiv 
	M_{\ss \rm P}^2R/D
	=\frac{3M_{\ss \rm P}^4}{4\alpha}
	\Biggl[ -1\pm\sqrt{1+\frac{8\alpha\Lambda_0}{3M_p^4}} \
	\Biggr].
\end{equation}
In case of a hierarchy~$|\alpha\Lambda_0| \! \ll \! M_{\ss \rm P}^4$,
the Einstein-Hilbert term balances out the bare 
cosmological constant term in the first branch, 
with the Gauss-Bonnet term providing a small correction, 
\begin{equation}
\Lambda_+\simeq
	\Lambda_0\left(1-\frac{2\alpha\Lambda_0}{3 M_{\ss \rm P}^4}\right) \, ,
\label{first branch}
\end{equation}
while in the second branch, reversely, 
the Einstein-Hilbert term balances out the Gauss-Bonnet term, 
while the bare cosmological constant only provides a small correction,
\begin{equation}
\Lambda_-\simeq -\frac{3M_{\ss \rm P}^4}{2\alpha}-\Lambda_0 \, .
\label{second branch}
\end{equation}
The existence of two branches of de Sitter solutions 
(or AdS solutions depending on the signs of $\alpha$ and $\Lambda_0$) in higher 
dimensional ($D\!\ge\!5$) 
Einstein-Gauss-Bonnet gravity  is well known in the literature. For instance, 
see Ref. \cite{Boulware:1985wk} for an early work. 
In the four dimensional limit of that solution the second branch in~(\ref{second branch})
is removed, and only the first branch in~(\ref{first branch}) remains as a solution.
However, in our setup, both of these branches remain
in four dimensional space-time as legitimate solutions
due to the rescaling in Eq.~(\ref{coupling}).

The question from the end of the Introduction section can be posed here in
a precise way: being in one of branches of the maximally symmetric solutions,
can we discriminate our theory from general relativity, at least at the 
level of  perturbation theory? To this end, we perturb the metric 
\begin{equation}
g_{\mu\nu} = \overline{g}_{\mu\nu} +  h_{\mu\nu} \, .
\end{equation}
where $\overline{g}_{\mu\nu}$ is the background (anti-)de Sitter metric.
A straightforward computation gives us the full equation of motion 
for linearized graviton evaluated in $D\!=\!4$,
\begin{widetext}
\begin{equation}
 \biggl( 1+ \frac{4\alpha}{3}\frac{\Lambda}{M_p^4} \biggr)
\biggl[ \nabla^\rho \nabla^\mu h_{\nu \rho} + \nabla_\nu \nabla_\rho h^{\mu \rho}
		 - \square {h^\mu}_\nu -  \nabla^\mu \nabla_\nu {h^\rho}_\rho
	+ \delta^{\mu}_{\nu} 
	\Bigl( \square {h^\rho}_\rho - \nabla_\rho \nabla_\sigma h^{\rho \sigma} \Bigr)
	+ \frac{\Lambda}{M_p^2}\Bigl( \delta^\mu_\nu {h^\rho}_\rho - 2 {h^\mu}_\nu \Bigr)
	\biggr] = 0.
\label{full pert eom}
\end{equation}
\end{widetext}
The  correction arising from Gauss-Bonnet 
term only appears in the overall factor of the equation of motion, while all 
terms in the brackets coincide with the ones from Einstein gravity.  
This result warrants two remarks. Firstly, the equation of motion being identical 
to the one of Einstein gravity implies that a graviton has only two degrees of freedom, 
which is consistent to what we expected from the beginning. 
Secondly, it implies that the effect of the Gauss-Bonnet action is only to shift
the Planck 
mass by a constant and thus its contribution to the linearized dynamics is trivial. 
However, the
sign of the overall factor in~(\ref{full pert eom}) would imply that the second 
branch~(\ref{second branch}) is unstable regardless of the sign~of~$\alpha$
(as noted in~\cite{Boulware:1985wk} for~$\alpha\!>\!0$), due to the overall
``wrong" sign in front of the linearized graviton action. This instability however
cannot indicate a spatially homogeneous decay since the only FLRW solutions
of~(\ref{EGB}) are the maximally symmetric de Sitter solutions. This is in contrast
to as the conformal anomaly ({\it e.g.}~\cite{Koksma:2008jn}), where the richer dynamics 
of the scale factor is attributable to the extra degrees of freedom.

From~(\ref{full pert eom}) we are
unable, to discriminate our theory from general relativity,
at the level of perturbation theory in a maximally symmetric 
space-time.
It is still possible though that this degeneracy is 
specific to the maximal symmetry of 
space-time, rather than of a more fundamental origin. Next we shall 
consider two less symmetric space-times: cosmological FLRW space-time and 
static spherically symmetric space-time. 
\\

{\bf FLRW Cosmology.}
~In order to study cosmology we consider the theory in~(\ref{EGB})
together with a scalar field, namely $S\!=\!S_{\rm \ss EH}\!+\!S_{\rm \ss GB}\!+\!S_{\phi}$,
where the scalar is canonical and minimally coupled to gravity,
\begin{equation}
S_{\phi}[g_{\mu\nu}, \phi]
	=\int \!d^{D\!}x \, \sqrt{-g} \,
	\biggl[ -\frac{1}{2} g^{\mu\nu} \partial_\mu \phi \, \partial_\nu \phi
		- V(\phi) \biggr],
\end{equation}
Assuming the FLRW ansatz~$ds^2\!=\!-dt^2+a^2d\boldsymbol{x}^2$, 
the Friedmann equations in~$D\!\to\!4$ limit read,
\begin{align}\label{fre2}
3M_{\ss \rm P}^2 H^2+6\alpha H^4 
	={}& \frac{1}{2}\dot{\phi}^2+V(\phi),
\\
- M_{\ss \rm P}^2\Gamma\dot{H}
	={}&\frac{1}{2}\dot{\phi}^2\,,
\end{align}
where we have defined a dimensionless 
parameter~$\Gamma\!\equiv\!1 + 4\alpha H^2/M_{\ss \rm P}^2 $.
The two Friedmann equations above are consistent with each other, 
provided that the scalar field satisfies the equation of 
motion~$\ddot{\phi}+3H\dot{\phi}+\frac{\partial V}{\partial\phi}=0.$ 
Therefore, the Bianchi identity  holds. 

One of the key observables in a FLRW universe is the transverse and traceless part of 
the metric fluctuation -- gravitational waves or tensor modes -- which we define as
\begin{equation}
g_{ij} = a^2 \bigl( \delta_{ij}+\gamma_{ij} \bigr) \, ,
\end{equation}
where~$\gamma_{ij}$ satisfies $\partial_i \gamma_{ij} \!=\! 0$ and~$\gamma_{ii} \!=\! 0$.
At the linear level these tensor modes are gauge invariant and decouple
from the vector and scalar modes due to
the spatial $SO(3)$ rotational symmetry.
Their equation of motion reads
\begin{align}
\ddot{\gamma}_{ij}
	+3 H \biggl( 1 - \frac{8\alpha \epsilon H^2}{3M_{\ss \rm P}^2\Gamma} \biggr)
		\dot{\gamma}_{ij}
	- c_{ \rm s}^2\frac{\partial^2\gamma_{ij}}{a^2}=0\,,
\end{align}
where~$\epsilon \!\equiv\!-\dot{H}/H^2 $, 
and~$c_{ \rm s}^2 \!\equiv\! 1 \!-\! 8\alpha \epsilon H^2/(M_{\ss \rm P}^2\Gamma)$, 
and an overall factor of $\Gamma$ has been omitted. 
Here again the $D\!\to\!4$ limit is well defined since the divergent factor 
in~(\ref{coupling}) is cancelled by the vanishing one in~(\ref{eomgb}).
The Gauss-Bonnet term modifies both the sound speed and the Hubble 
friction term compared to the general relativity limit~$\alpha\!=\!0$. During the 
early universe, the inflationary epoch for instance,
when~$H^2/M_{\rm \ss P}^2$ is not as small as nowadays,  
we would expect some non-trivial observational effects, given a 
reasonably sized $\alpha$. At late times, however,~$H^2/M_{\rm \ss P}^2$ 
is tiny and we thus expect the predictions from gravitational 
waves sector are consistent with all current astrophysical and 
cosmological observations, including the multi-messenger gravitational 
waves detection of binary neutron star merger~\cite{TheLIGOScientific:2017qsa}.

Another important observable in the FLRW universe is the scalar cosmological 
perturbation, which is essentially due to the single scalar field~$\phi$ in the matter 
sector and the scalar polarization of metric fluctuation it induces.
We  define the scalar perturbation on the metric as follows, 
\begin{align}
&
g_{00} = - ( 1+2\chi ) \, ,
\qquad \quad
g_{0i}=\partial_i\beta \, ,
\nonumber \\
&
g_{ij} = a^2 e^{2\zeta} (\delta_{ij}+\partial_i\partial_jE ) \, .
\end{align}
We have to perturb the scalar field as well, 
\begin{equation}
\phi(t,\textbf{x})=\phi(t)+\delta\phi(t,\textbf{x}).
\end{equation}
Noted that the theory possesses full space-time diffeomorphisms, and therefore we 
can safely remove $\delta\phi$ and $\partial_i\partial_jE$ by performing the following
coordinate transformation,
\begin{equation}
t\to t+\xi^0 \, ,
\qquad 
x^i\to x^i+\partial_i\xi \, ,
\end{equation}
given proper function of $\xi^0$ and $\xi$. Among the rest of three scalar 
variables, $\chi$, $\beta$ and $\zeta$, we find $\chi$ and $\beta$ are non-dynamical. 
We can eliminate these two non-dynamical modes by solving the $(00)$ and $(0i)$
components of Einstein equations, {\it i.e.} solving the Hamiltonian constraint 
and momentum constraint equations.  Doing so results in the equation of motion for the scalar mode,
\begin{equation}
\ddot{\zeta}
	+3H \biggl( 1+\frac{\eta}{3}
			- \frac{8\alpha \epsilon H^2}{3M_{\ss \rm P}^2\Gamma} \biggr)
				\dot{\zeta}
	-\frac{\partial^2\zeta}{a^2}=0 \, ,
\end{equation}
where~$\eta \!\equiv\! \dot{\epsilon}/H\epsilon$,
and again  the overall factor~$\epsilon \Gamma$ has been omitted.
We see that the Hubble friction term of the scalar mode is modified,
while its sound speed is unity. The sound speed of scalar mode is 
generally different from the one of gravitational waves. However, this deviation is 
tiny in the late universe, as it is proportional to~$H^2/M_{\ss \rm P}^2$.

The tensor and the scalar perturbations are all the physical degrees of freedom in
the theory given by~$S_{\ss \rm EH} \!+\! S_{\ss \rm GB} \!+\! S_{\phi}$, as is expected since
the Gauss-Bonnet action does not give rise to any additional degrees of freedom
when added to the Einstein-Hilbert one in any number of space-time dimensions.
Therefore, no vector modes are expected, which we have confirmed by checking 
that they are all eliminated by solving for the momentum constraint equations.
\\

{\bf Static Spherically Symmetric Solution.}
We now derive the static spherically symmetric solution
for the theory given by~$S_{\ss \rm EH}\!+\!S_{\ss \rm GB}$, 
with the vanishing bare cosmological constant.
It is clear from the onset that the Schwarzschild metric does not solve the vacuum
Einstein's equations, on the account that the Riemann tensor, 
which appears explicitly 
in~(\ref{eomgb}), does not vanish.
As we shall see shortly, vacuum equations with the Gauss-Bonnet term 
allow for solutions free from the
the singularity issue of general relativity. 
The solutions for a static and spherically symmetric ansatz in an arbitrary number of 
dimensions~$D\!\ge\!5$,
\begin{equation}\label{sphansatz}
ds^2 = -e^{2\omega}dt^2
	+e^{2\lambda}dr^2
	+r^2d\Omega_{D-2}^2
\end{equation}
were already found in Ref.~\cite{Boulware:1985wk}. 
These are extended to~$D\!=\!4$ solutions of our theory in~(\ref{EGB})
by making the rescaling~(\ref{coupling}),
and then taking the limit~$D\!\rightarrow\!4$,
\begin{align}
&
-g_{00} \! = \!e^{2\omega} = \! e^{-2\lambda} \!
\label{sch-de}
\\
&	\hspace{1.2cm}
	= 1 + \frac{r^2}{32 \pi \alpha G}
	\Biggl[ 1\pm \biggr( \!1 \!+\! \frac{128 \pi\alpha G^2 M}{r^3} \biggr)^{\!\! 1/2 \,} \!
	\Biggr] \, .
\nonumber 
\end{align}
Here instead of the reduced Planck mass we give results in terms
of more customary Newton's constant,~$G \!=\! 1/(8\pi M_{\ss \rm P}^2)$,
and  $M$ is a test point mass.
There are two branches of solutions if~$\alpha\!>\!0$. 
However, if $\alpha\!<\!0$, there is no real solution at short radial distances
for which~$r^3\!<\!-128\pi \alpha G^2 M$. 
The absence of real solutions at short distances implies the static spherically symmetric 
ansatz in Eq.~(\ref{sphansatz}) is not a good assumption, and
probably we need a more general ansatz to find real solutions. 
This problem is beyond the scope of the {\it Letter} at hand and we will leave it for 
future investigations. In this section we focus on the case $\alpha>0$. 

\begin{figure}
\begin{center}
\includegraphics[width=0.45\textwidth]{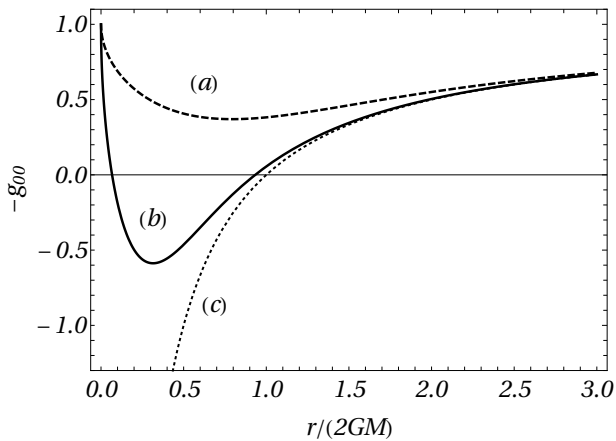}
\end{center}
\caption{ Radial dependence of gravitational potential~$g_{00}$ in the four-dimensional 
Einstein-Gauss-Bonnet gravity in 
cases~$(a)$~$\alpha\!=\!M^2 G/(4\pi) $~($M\!=\!M_*/2$, dashed line)
and~$(b)$~$\alpha\!=\!M^2 G/(64\pi)$~($M\!=\!2M_*$, full line), and in~$(c)$~general relativity ($\alpha\!=\!0$, dotted line).
}
\label{gpot}
\end{figure}

At large distances the two branches behave asymptotically as
\begin{equation}\label{asymt}
-g_{00} 
	\stackrel{r \rightarrow\infty}{\sim} 1-\frac{2GM}{r}
\quad \text{or} \quad
	1+\frac{r^2}{16\pi\alpha G}+\frac{2GM}{r} \, ,
\end{equation}
{\it i.e.} they reduce to a Schwarzschild solution with positive gravitational mass,
or to a Schwarzschild-de Sitter solution with negative gravitational  mass, respectively.
We are more interested in the first branch, 
the one with minus sign inside of brackets in Eq.~(\ref{sch-de}), where we have 
asymptotic Schwarzschild metric at large distance.   Note that even though this branch is a vacuum solution,
the Ricci scalar does not vanish due to the contributions from the Gauss-Bonnet
term to the vacuum Einstein's equations. 
The physical properties of this branch differ depending whether the mass~$M$
is larger or smaller than the critical mass given by
\begin{equation}
M_{*} = \sqrt{\frac{16\pi \alpha}{G}} \, ,
\end{equation}
and in Fig.~\ref{gpot} we plot the radial dependence of~$g_{00}$ to illustrate it for
(a)~$M\!<\! M_*$, and (b)~$M\!>\! M_*$.
In both cases the gravitational potential has a minimum, and gravity is therefore
attractive to the right of the minimum, and repulsive to the left of it. What distinguishes
the two cases is that in the first case the gravitational potential is always positive,
and there are no horizons that form, and hence no black hole solutions, while in the second 
case the gravitational potential crosses zero at two points defining two horizons,
\begin{equation}
r^{\ss H}_\pm = GM \Biggl[ 1 \pm\sqrt{1 - \frac{16 \pi \alpha}{GM^2}} \ \Biggr] \, .
\end{equation}
The horizon at~$r^{\ss H}_+$ is the event horizon of a black hole, 
which envelops a white hole with the event horizon at~$r^{\ss H}_-$.
We expect the gravitational collapse comes to a halt when the size of system reaches 
the one corresponding to the bottom of the gravitational potential for a collapsing dust 
model. In a realistic stellar collapse, the gravitational collapse ceases at somewhere 
between the bottom of the potential and the event horizon of a black hole due to the stellar 
internal pressure.

Another important property is the resolution for the singularity problem. At short distances $r\to0$, the gravitational potential approaches a finite value $-g_{00} {\to} 1$, while the curvature invariant $R\propto r^{-3/2}$ diverges at short distance limit (so does the  gravitational force). Nevertheless, the gravitational force is repulsive at short distance and thus an infalling particle never reaches $r=0$ point. In this sense, our theory is practically free from singularity problem. This is in contrast to Einstein's general relativity, where an infalling particle will eventually hit the singularity and effective theory breaks down.
\\

{\bf Conclusion and Discussion.}
~The Gauss-Bonnet action does not contribute to the dynamics
of the four dimensional space-time, 
as its contribution to Einstein's equation vanishes identically in~$D\!=\!4$ space-time
dimensions. We multiply the Gauss-Bonnet action by a factor of $1/(D\!-\!4)$ 
to compensate for this and to produce a finite non-vanishing contribution to
Einstein's equations in~$D\!=\!4$.	Thus the Gauss-Bonnet action becomes a 
non-trivial ghost-free extension of the Einstein-Hilbert action.
It should be noted that the limit $D\!\to\!4$ has to be taken in the continuous
sense, at the level of the equations of motion, rather than 
at the level of the action. It is in general possible to take this limit~\cite{Mardones:1990qc,Torii:2008ru}, however, in practice it should be taken with due care.

In the several examples we presented we were able to take the continuous~$D\!\to\!4$
limit in a natural and straightforward way due to to the assumed symmetry between a 
number coordinates of the space-time solving Einstein's equations. It is not obvious though,
that this works in less symmetric space-times. In our prescription 
the additional dimensions have no physical meaning, and only serve to define the limit.
Therefore, in practice one can extend the dimensionality of space-time in a way that the 
symmetries between coordinates are restricted to the fiducial dimensions and one
physical spatial dimension. Thus the limit is finite and well defined. However, 
additional important insight can be obtained from different formulations of the theory,
other than the tensor formalism we utilize here.

In~\cite{Torii:2008ru} the Gauss-Bonnet term was examined in the $D\!=\!d\!+\!1$ ADM
decomposition. Upon rescaling~(\ref{coupling}) of the coupling constant we employed,
one can read of from the canonical equations that the 
dynamical~$(ij)$ Einstein's equations are manifestly finite in~$D\!=\!4$
as the singular term explicitly cancels (Eq. (76) from~\cite{Torii:2008ru}). This is 
however not manifest in the constraint sector, where the limit has to be taken carefully.
Nevertheless, the precise prescription of the limit in the constraint sector cannot be
of physical concern, as the constraint sector has to be chosen by hand 
anyway,~{\it i.e.} we have to specify gauge conditions.

Even more insight is provided by formulating the Gauss-Bonnet action  in terms of 
differential forms. In~$D\!=\!4$ it is just an Euler 
density,~$S_{\ss \rm GB}\!\sim\!\int\epsilon_{a_1...a_4}R^{a_1a_2}\wedge R^{a_3a_4}$,
which is just a total derivative. In~$D\!>\!4$ the Gauss-Bonnet term 
reads~$S_{\ss \rm GB} \! \sim\! \int\epsilon_{a_1...a_D} R^{a_1a_2}\wedge R^{a_3a_4}\left(\wedge e^a\right)^{D-4}$, which is an exterior product of a total derivative 
and a~$(D\!-\!4)$-form. Taking the variation with respect to the {\it vielbein} gives rise to 
the vanishing factor $(D\!-\!4)$ which is precisely cancelled out by the singular factor in the
 coupling constant rescaling eq. (\ref{coupling}), and we thus expect that all components of 
the Einstein equation are regular. 
\\

Similar idea to the one presented here
has been considered before motivated by the study
of quantum corrections arising from integrating out matter 
fields~\cite{Tomozawa:2011gp,Cognola:2013fva}.
The perspective that we take is that the Gauss-Bonnet action should be considered
a classical modified gravity theory, defined by
a modified action principle, rather than a one-loop perturbative correction.
In that sense it is on an equal footing with general relativity.

The Gauss-Bonnet extension to Einstein's gravity presented here satisfies 
the criteria of Lovelock's  theorem.
In general it leads to very different 
phenomenologies. For the spherically symmetric static solution it predicts singularity
resolution. Generally there are two event horizons for a spherical 
static solution in vacuum. The interior horizon is an event horizon of a white hole, 
enveloped by the event horizon of a black hole, so a gravitational collapse ceases 
with a typical length scale somewhere in between. Cosmological applications of
our theory imply a modified dispersion relation for the tensor modes. This has 
potential observational relevance as it provides a possibility of the parametric 
resonance, and the production of gravitational waves during the reheating epoch.

We expect a similar prescription presented here to apply to higher 
order Lovelock invariants. These are 
of sub-sub-leading effects in Einstein equation in a weak field limit, compared to 
the Einstein-Hilbert term and the finite Gauss-Bonnet term. Therefore, this class of 
theories bypasses the conclusions of Lovelock's theorem on the
account of modifying the action principle, and challenges the distinctive role of 
general relativity as the unique non-linear theory describing gravitational interactions in 
the four dimensional space-time. 
\\

{\bf Acknowledgments.}
We are grateful to Sergio Zerbini for drawing our attention to 
Refs.~\cite{Tomozawa:2011gp,Cognola:2013fva}.
D.~G.~is grateful to the Institute of Theoretical Physics of the University 
of Warsaw for the hospitality during the initial stages of the project.
C.~L. would like to thank Z. Lalak and R. Brandenberger for the useful discussions. D.~G. is supported by the Fonds de la Recherche 
Scientifique -- FNRS under Grant IISN 4.4517.08 -- Theory of fundamental
interactions. 
The work of C.~L.~is carried out under
POLONEZ programme of Polish National Science Centre, 
No. UMO-2016/23/P/ST2/04240, which has received funding from the 
European Union's Horizon 2020 research and innovation programme under 
the Marie Sk\l{}odowska-Curie grant agreement No. 665778.

\end{document}